\RequirePackage{fix-cm}

\documentclass[twocolumn]{svjour3}          

\smartqed  

\usepackage{graphicx}
\usepackage{siunitx}
\usepackage{float}
\newcommand{\Mg}{\textsuperscript{24}\text{Mg}\textsuperscript{+}\,}
\newcommand{\MgH}{\textsuperscript{24}\text{MgH}\textsuperscript{+}\,}
\newcommand{\Ba}{\textsuperscript{138}\text{Ba}\textsuperscript{+}\,}
\journalname{Applied Physics B}

\begin{document}

\title{Deterministic delivery of externally cold and precisely positioned single molecular ions
}


\author{G. Leschhorn        \and
        S. Kahra            \and
        T. Schaetz 
}


\institute{G. Leschhorn \and
           S. Kahra     \and
           T. Schaetz   \at
              Max-Planck-Institut for Quantum Optics \\
              Hans-Kopfermann-Str.1\\
              85748 Garching, Germany\\
              Tel.: +49 89-32905-199\\
              Fax: +49 89-32905-200\\
              \email{tobias.schaetz@mpq.mpg.de}          
\and
					T. Schaetz	\at
							Albert-Ludwigs-Universit\"at Freiburg\\
							Hermann-Herder-Str. 3a\\
							79104 Freiburg, Germany           }

\date{}

\maketitle

\begin{abstract}
We present the preparation and deterministic delivery of a selectable number of externally cold molecular ions. A laser cooled ensemble of \Mg ions subsequently confined in several linear Paul traps inter-connected via a quadrupole guide serves as a cold bath for a single or up to a few hundred molecular ions. Sympathetic cooling embeds the molecular ions in the crystalline structure. \MgH ions, that serve as a model system for a large variety of other possible molecular ions, are cooled down close to the Doppler limit and are positioned with an accuracy of one micrometer. After the production process, severely compromising the vacuum conditions, the molecular ion is efficiently transfered into nearly background-free environment. The transfer of a molecular ion between different traps as well as the control of the molecular ions in the traps is demonstrated. Schemes, optimized for the transfer of a specific number of ions, are realized and their efficiencies are evaluated. This versatile source applicable for broad charge-to-mass ratios of externally cold and precisely positioned molecular ions can serve as a container-free target preparation device well suited for diffraction or spectroscopic measurements on individual molecular ions at high repetition rates (kHz).

\end{abstract}

\section{Introduction}
\label{introduction}
Following different approaches, sources of cold molecules have substantially advanced in resent years. Trapping \cite{Bethlem2000} \cite{junglen04b}, deceleration \cite{Bethlem1999} and filtering \cite{rangwala03} of molecules and molecular ions are only a few examples. The huge amount of vibrational and rotational states and the related effort required to realize closed transitions hinders the effective use of direct laser cooling. Alternative approaches, for example sympathetic cooling of molecular ions within heterogeneous Coulomb crystals, have been demonstrated \cite{molhave00} \cite{Roth2005}. Ranging from fundamental physics to a large variety of applications in physics and chemistry, trapped molecular ions are nearly ideal objects to study interactions of single, well prepared and localized particles.

One application requiring single, precisely positioned molecular ions is coherent hard X-ray diffraction for structural determination with atomic resolution and time-resolved pump-probe experiments using short X-ray pulses \cite{neutze00,chapman06}. The weak interaction of X-rays with matter requires the Bragg diffraction from a solid state crystalline sample to achieve a sufficient signal gain \cite{webster02}. Currently, the prerequisite of crystallization is the most severe constraint of these techniques, because many biological molecules are difficult or impossible to crystallize \cite{henderson95}. The proposal \cite{neutze00} of diffracting the short pulses of a free-electron laser on single molecules eliminates the mentioned constraints and even proposes a solution to the problem that the huge radiation dose alters the nuclear skeleton or even destroys the sample-molecule. However, the required amount of photons to image a single molecule ($10^{13}$ per pulse ) has to be bunched within a sufficiently short exposure allowing to record useful structural information before the radiation degrades the sample-molecule detrimentally. In order to reach this high photon flux a focus of approximately 100\,nm diameter is required \cite{neutze00,young10}. The realization of a micrometer sized beam of X-rays has already been reported \cite{seibert11}. Identically initialized, single trapped molecular ions provide the required localization and represent a container-free target at minimal background. Trapped ions provide excellent control over the external degrees of freedom down to the motional ground state within the confining potential \cite{barrett03}. A manipulation of the internal degrees can be achieved with state-of-the-art methods like buffer gas cooling \cite{gerlich09}. Replacing of the target with the pulse repetition rate of the X-ray source on the 0.1-1\,kHz level should be feasible. This work reports on a source for generic molecular ions, represented by \MgH, sympathetically cooled into a Coulomb crystal of directly laser cooled atomic \Mg-ions and transfered into ultra-high vacuum with an efficiency close to unity.

\section{Experimental setup}
\label{the experimental setup}
\begin{figure*}
	\centering
		\includegraphics[width=0.9\textwidth]{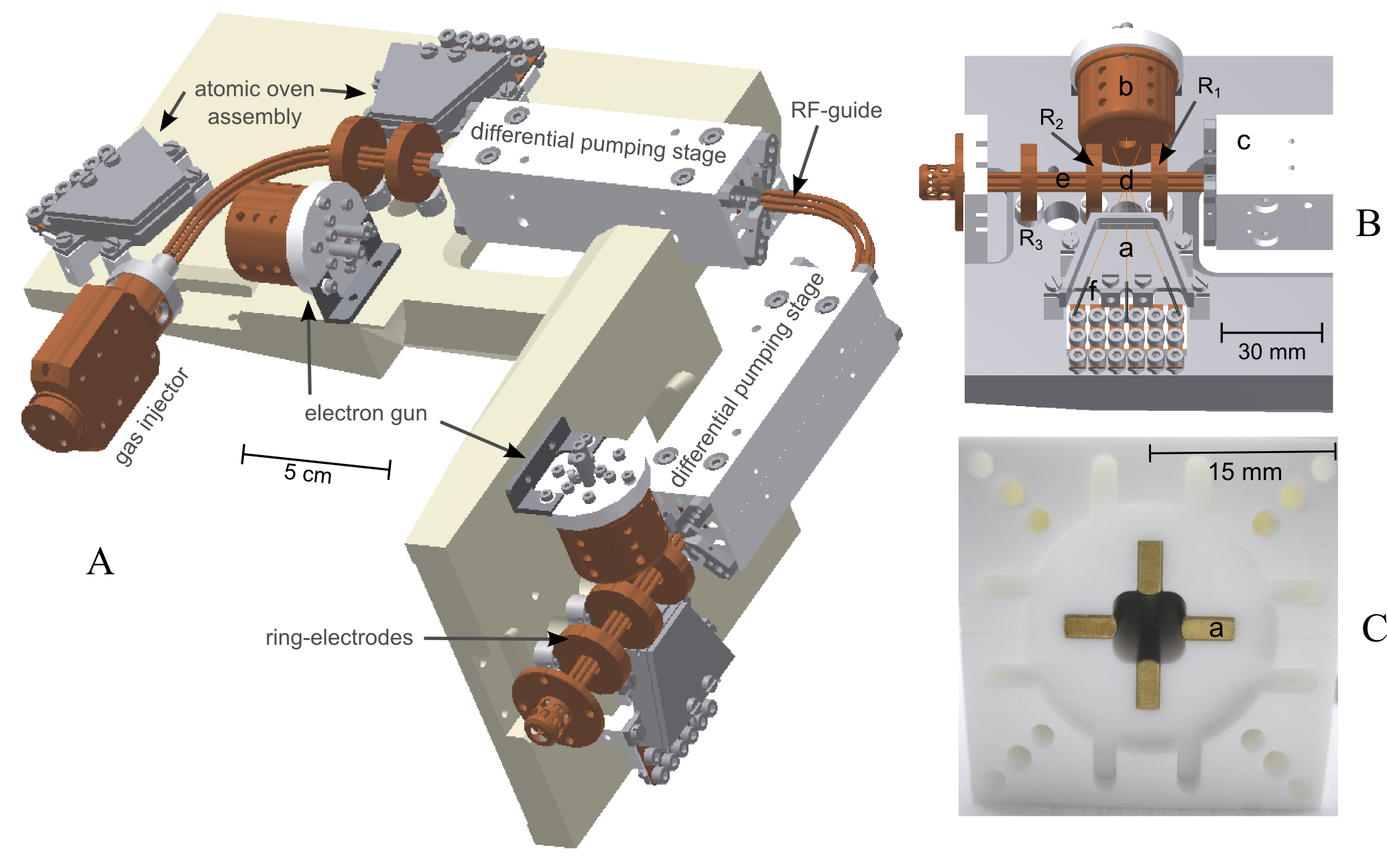}
		\caption{Overview of the most important parts of the guide for (molecular) ions and optional traps. Various elements have been removed for the sake of clarity. (A) shows the cryogenic gas injector and the double bent RF-guide. Two differential pumping stages separate three different chambers, that can be pumped individually. Five ring electrodes allow to form three trapping regions, where the confined particles can be observed by detecting the fluorescence light of laser cooled \Mg or \Ba (see figure \ref{fig:ionenbild}). Three assemblies of atomic ovens and two electron guns are positioned along the guide. (B) gives a close-up of the traps 2 (d) and 3 (e) at the end of the guide. Neutral \textsuperscript{24}\text{Mg} atoms can be evaporated into the trap by resistively heating one of the Mg-filled tantalum tubes (f) of the atomic oven assembly (a) and can be ionized either by photoionization or by electron bombardment ionization using electron gun (b). (C) shows a close-up on the front side of one of the 10\,cm long ceramic tubes, that serve as differential pumping stages. The clover leaf shaped profile leaves a 0.5\,mm gap to the RF-electrodes and minimizes the open cross section between the different chambers. Voltages on the four gold plated copper inlets (a) can either serve as compensation voltages to counteract stray fields or form drift tubes to bunch the accelerated and guided ions. Additionally, they do not charge up like ceramics, that would hinder a smooth transfer of ions.}
	\label{fig:apparat}
\end{figure*}

An overview of the mechanical layout is presented in figure \ref{fig:apparat}. The ultra-high vacuum (UHV) chamber contains a linear quadrupole guide \cite{Paul90} providing a two dimensional confinement for ions. Direct current (DC) voltages on ring-electrodes enclosing the guide can either be used to accelerate and decelerate ions along the guide axis or allow for additional axial confinement required to turn three sections of the quadrupole into linear Paul-traps. Four cylindrical, gold plated copper rods of 2\,mm diameter are arranged in quadrupole configuration, such that the minimal distance between the guide center and the surface of the rods amounts to $\text{r}_0=1.12$\,mm. The 49\,cm long guide is bent twice by $90^\circ$ with a radius of curvature of 35.5\,mm for a different application. This will allow to investigate the efficiency of the guide to transfer cold neutral polar molecules using the Stark effect \cite{rangwala03} in a radio-frequency (RF) field and to explore their interactions with trapped molecular ions \cite{willitsch08}. Concerning the experiments presented in this paper, the double bend enhances the efficiency of extracting the injected gas, required for the generation of the molecular ions. Radial confinement along the quadrupole is achieved by applying a RF-voltage on each pair of opposing electrodes with a mutual phase difference of $\pi$ and a frequency of $\Omega_\text{RF}=2\pi\cdot6.8$\,MHz. To allow for ions with small charge-to-mass ratios (Z/m), stable trapping conditions require a high resonance frequency of the used helical RF-enhancement resonator. Therefore, parasitic capacities, caused by e.g. dielectrics, are reduced to a minimum by the chosen electrical layout. Due to the skin effect, the RF-current flows only in a thin surface layer in copper of approximately 25\,\micro m . A 3\,\micro m thick gold coating protects the surface of the copper electrodes from oxidation during the assembly and large contacting clamps maximize the electric and thermal contact. This leads to a high enhancement factor for RF-voltages of Q=730 (loaded) and thus supports the high voltages needed for a stiff confinement of (molecular-) ions with small Z/m and the envisioned experiments with neutral molecules. 

Pairs of 5\,mm thick copper ring electrodes (outer diameter: 24\,mm, inner diameter: 8\,mm), centered along the axis of the guide, create regions of 3D confinement (refer to figure \ref{fig:apparat}) and additionally serve as electrodes for the acceleration and controlled transfer of the ions. A DC-voltage applied to each pair of these electrodes, spaced by 15\,mm, generates a nearly harmonic potential in the central trapping region leading to static confinement along the guide axis. The ring electrodes have been carefully designed to account for mechanical stability, optical access to the axis of the guide and the shielding effects caused by the RF electrodes (see also figure \ref{fig:beschleunigung}). The ion trap, completed by the two ring electrodes after the first bend of the guide, is labeled trap 1 in the following. At the end of the guide, three ring electrodes can create two trapping regions, named trap 2 and 3. The middle electrode is shared, resulting in a double well axial potential. Optionally, the traps can be combined to a larger one. In total, up to three traps can be used. Typically, for \Mg, the measured radial frequency in trap 2 amounts to 410-540\,kHz, which corresponds to a root-mean square RF-voltage on the electrodes of 34-46\,V. The frequently used configuration of 210\,V and 100\,V on the two ring electrodes of trap 2 leads to a typical axial trap frequency of about $2\pi\cdot30\,\text{kHz}$. The asymmetry of the applied voltages leads to a trap minimum at an axial position where micro-motion \cite{Paul90} can be compensated best.

\begin{figure}
	\centering
		\includegraphics[width=0.5\textwidth]{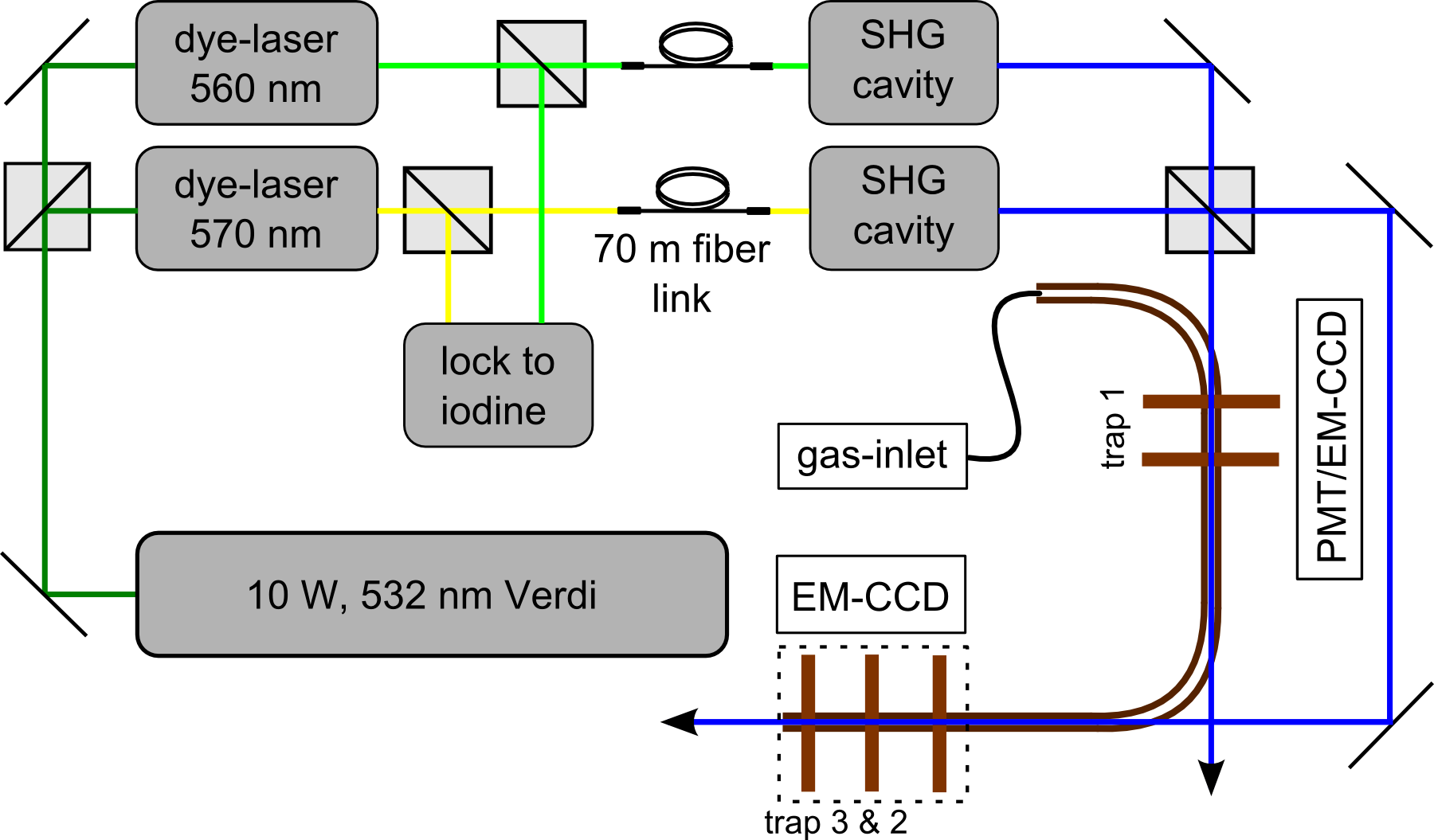}
		\caption{Schematic overview of the experimental setup including the laser part for photoionization of neutral magnesium and Doppler cooling of \Mg. Two second harmonic generation ring cavities are used for frequency doubling the output frequencies of two dye-lasers, which are stabilized and locked to two Doppler-free iodine spectroscopy setups. The resulting 280\,nm light for cooling of \Mg and the 285\,nm light for photoionization of neutral magnesium are overlapped using a polarizing beam splitter and enter the quadrupole (see figure \ref{fig:apparat}) in the vacuum chamber along its symmetry axis via two laser ports. Ring electrodes form three trapping regions, where the confined particles can be observed by detecting the fluorescence light of laser cooled \Mg. The observation in trap 1 is achieved by a photo-multiplier tube or a UV sensitive electron multiplying CCD camera, respectively. Ions in trap 2 and 3 can be imaged on a CCD camera.}
	\label{fig:setup}
\end{figure}

Trap 1 and 2 can be selectively loaded with all stable isotopes of \text{Mg}\textsuperscript{+} and \text{Ba}\textsuperscript{+} ions by applying photoionization schemes \cite{kjaegaard00,rotter08,Bapapier} on atoms out of a thermal beam, evaporated into the trapping region by heating one of three 0.8\,mm thick metal wires inside tantalum tubes. One of the housed stacks of three atomic ovens (f) can be seen without cover in figure \ref{fig:apparat} B labeled (a). Photoionization is far more efficient and less disturbant compared to electron-impact ionization \cite{madsen00}. A smaller temperature of the resistively heated atomic ovens is required. As a consequence, the deposition of magnesium or barium on the trap electrodes and the influence of the related contact potentials is reduced. The smooth transfer of molecular ions along the guide and its efficiency is enhanced. Furthermore, charging of dielectrica by the electron gun is avoided. For photoionization of magnesium a resonant two-photon process is used. One photon with a wavelength near 285\,nm excites the neutral magnesium atom from its ground state $\text{3s}^2\,\text{S}_0$ to the $\text{3s3p}\,\text{P}_1$ state. A second photon (285\,nm) or a 280\,nm photon of the Doppler cooling laser for \Mg (see figure \ref{fig:setup}) is sufficient to reach the continuum. All translational degrees of freedom of the \Mg ions can be laser cooled down to the Doppler limit ($\approx$ 1\,mK) driving the $\text{3s}\,\text{S}_{1/2}\leftrightarrow \text{3p}\,\text{P}_{3/2}$ transition with a natural linewidth of $\Gamma\approx 2\pi \cdot43$\,MHz. If the kinetic energy of the ions at a given density falls below a critical value, a phase transition into a crystalline structure, known as a Coulomb-or Wigner crystal, occurs \cite{diedrich87b}\cite{wineland87}. The lattice parameter of such a crystal is of the order of a few \micro m. The localization of an ion cooled to the ground-state of the quantized harmonic trapping potential is ultimately limited by the width of the ground state wavefunction ($\approx$20\,nm for \Mg at $\omega_{sec}\approx2\pi\cdot500\,\text{kHz}$). Typically, a Doppler cooled ion can be localized to better than 1\,\micro m, currently limited by the imaging properties of the used objective.

The laser setup for photoionization of neutral magnesium and laser cooling of \Mg is shown in figure \ref{fig:setup}. Two dye-lasers are pumped each by 3.5 W of a frequency doubled Nd:YAG laser and provide a typical output power of 500 mW at 570\,nm and 250\,mW at 560\,nm. UV-light at the required frequencies for laser cooling and photoionization is created by two external second-harmonic generation (SHG) ring resonators incorporating non-linear BBO crystals \cite{friedenauer06}. The enhancement cavities and the trap apparatus share the same optical table. The connection to the lasers, that are located in a different laboratory, is achieved using two 70\,m long single mode optical fibers. The dye-lasers are frequency locked to suitable iodine transitions, using two different Doppler-free spectroscopy setups (Absorption spectroscopy \cite{smith71} for the 560\,nm and polarization spectroscopy \cite{wieman76} for the 570\,nm light). A much more compact setup can be implemented using an all solid state turn key laser system described in \cite{friedenauer06}.  

The fluorescence light of ions confined in trap 1 can either be observed by an imaging system consisting of an air-spaced, two lens condensor (magnification: 10$\times$) followed by an electron multiplying CCD camera or by a photomultiplier (see figure \ref{fig:setup}). The camera is used for applications, that require spatial resolution of individual ions while the photomultiplier provides a spatially integrated fluorescence signal, sufficient for a controlled loading of the trap. The magnification factor of the imaging system shared by trap 2 and 3 can be choosen to be 10 by using a three lens condenser or 50 by using a four lens near diffraction limited objective. Fluorescence light from these traps can be observed with a EM-CCD camera.

The apparatus offers several options for the creation of molecular ions of choice, discussed in greater detail in section \ref{production}. A cryogenic inlet, similar to the one in \cite{rangwala03}, can inject gas into the vacuum vessel via a pulsed valve. The gas pulses (few ms duration) pass a copper capillary, optimized for thermal contact to a liquid nitrogen bath, allowing for internally cold molecules, but can be additionally heated to evaporate a potential blockage and to carefully adjust the temperature of the gas, if required. After passing the copper-injection tube the molecules enter the chamber through a ceramic nozzle. A gap of 0.5\,mm between the front surface of the quadrupole and the nozzle mitigates the heat transfer. Right behind the injection system, an electron gun can ionize the neutral molecules in the center of the guide. The whole injection system can be replaced by an Electrospray Ionization source (ESI), based on soft protonation methods \cite{fenn89}.

Ultra-high vacuum of the order of $10^{-10}\,\text{mbar}$ is required for experiments relying on trapped and crystallized ions, to minimize reactions and collisions with the background gas. As described in reference \cite{ostendorf06}, the concept of using a quadrupole guide to transfer ions from the preparation zone to a spatially separated experimental zone allows for the required UHV, despite the comparatively high pressure ($\sim10^{-3}-10^{-6}$\,mbar), related to the efficient creation of molecular ions. Additionally, a buffer gas cooling stage \cite{gerlich09}, to address the molecule's internal degrees of freedom, can be implemented if required by the experiment. Currently, the quadrupole, starting at the gas inlet nozzle, guides the ions through three chambers, that are separated by two 10\,cm long ceramic tubes serving as differential pumping stages with an open cross section, that adapts to the quadrupole (see figure \ref{fig:apparat} C). Care has been taken to avoid uncontrolled charge ups, for example, on insulators near the guide, severely reducing the efficiency or even hindering the guiding of ions along the quadrupole. Therefore the ceramic parts of the tubes, that would remain in line of sight for the ions, have been replaced by gold plated copper bars. A voltage can be applied to these bars to enhance the transfer process. For the experiments presented in the next section, the copper bars are grounded. The differential pumping stages between individually pumped chambers allows to maintain a pressure difference of 4 orders of magnitude between the first and the last chamber, typically kept in the few $10^{-10}$\,mbar regime.

\begin{figure*}
	\centering
		\includegraphics[width=1\textwidth]{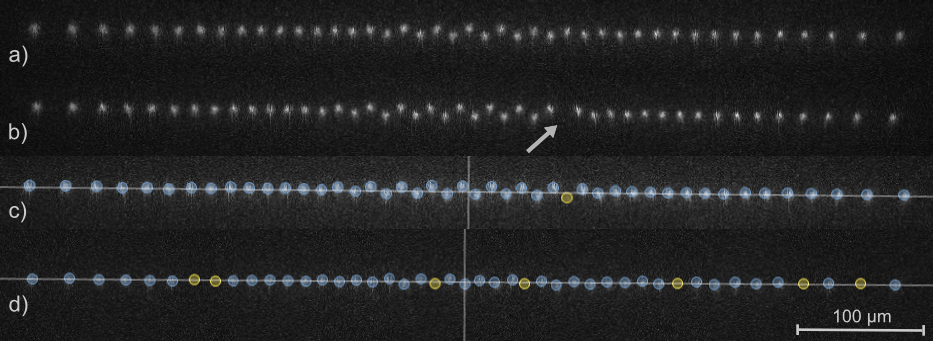}
		\caption{Two-dimensional fluorescence CCD images of crystalline ion structures confined in trap 2 (see fig. \ref{fig:apparat} B) under UHV conditions, using ten times magnification to image the whole ion crystal. Crystal a)  contains only \Mg ions and is referred to as a pure crystal. Crystal b) contains exactly one molecular \MgH ion indicated by the arrow. In c) the horizontal axis and the vertical symmetry axis of the crystal in b) is given by white lines. After the identification of the bright spots, the position of the atomic \Mg-ions is marked by blue circles. The symmetry of the Coulomb crystal is exploited to identify the position of the non-fluorescing molecular ion (\MgH) marked by a yellow circle with an accuracy of approximately one micrometer. The ion crystal d) demonstrates the preparation of 7 molecular ions. The number of sympathetically coolable and crystallizable molecular ions is determined by the cooling rates mediated via the atomic ions. One cooling ion can be sufficient to sympathetically cool and crystallize $\sim10$ molecular ions.}
	\label{fig:ionenbild}
\end{figure*}

\section{Results}
In this section, deterministic delivery and preparation methods for molecular ions and their integration and sympathetic cooling embedded in the crystalline structure of directly laser cooled atomic ions are described. After its creation, the molecular ion is transfered into an isolated region. For a single ion the efficiency of the transfer process is close to unity and fast ($\ll$ms). For the transfer of many individual molecular and atomic ions, three schemes suited for different applications are described. The deterministic delivery of a selectable number of externally cold molecular ions, allowing for repetition rates up to kHz, is persecuted and an accuracy for spatial positioning of a micrometer is demonstrated.

\subsection{Production}
\label{production}
The apparatus offers several schemes to create a molecular ion (see also section \ref{the experimental setup}) and to embed it on a lattice site of an ion crystal. Electron bombardment ionization of neutral molecules out of an effusive beam is possible using the electron gun, that is located behind the cryogenic gas inlet (see figure \ref{fig:apparat} (A)). On the one hand, a wide mass range of molecular ions can be produced using this method. On the other hand, the molecule, even if originally internally cold, will end up in a highly excited, dissociative state or the electron-impact might lead to an immediate dissociation. Optionally, an already realized post-filtered ESI setup can substitute the gas inlet stage and serve as a soft ionization device for larger molecules of selectable charge-to-mass ratio. To investigate the control and transfer capabilities of the setup, the production of the molecular ions is realized by the photochemical reaction \cite{molhave00}

\begin{equation}
	^{24}Mg^++H_2\rightarrow\, ^{24}MgH^++H.
	\label{reaction}
\end{equation}

A collision of an electronically excited \Mg ion with a hydrogen molecule leads to the formation of an magnesiumhydrid ion and an hydrogen atom. The \MgH ion serves as a model system to calibrate the setup for almost generic molecular ion species having a suitable charge-to-mass ratio such that they can be sympathetically cooled exploiting directly a laser cooled \Mg or \Ba crystal as a heat sink \cite{larson86}.

A reliable preparation of an isotopically clean (here \Mg) atomic crystal is the initial step. Different magnesium isotopes (\textsuperscript{25}\text{Mg}\textsuperscript{+}\,,\textsuperscript{26}\text{Mg}\textsuperscript{+}), as well as accidentally ionized residual gas particles, appear as dark lattice sites within the crystal (see figure \ref{fig:ionenbild} b) and c)) and are not distinguishable from the molecular ions. To keep the contamination negligible, the isotope selective two photon photoionization loading scheme, discussed in section \ref{the experimental setup}, is sufficient for the current needs. If the experiment does not require a fast or even continuous reloading, an optional cleaning of the ion crystal can be implemented. The contaminant ion's kinetic energy can be increased above the trap depth, by resonantly driving its specific secular motion \cite{Drewsen04}. An arbitrary waveform generator produces the sum of the radial secular frequencies ($\propto \text{U}_\text{RF}\cdot\text{Z/m}$) of particles with masses m=25\,u ($2\pi\cdot480\,\text{kHz}$) and m=26\,u ($2\pi\cdot462\,\text{kHz}$) and the resulting beat signal is applied to a copper wire parallel to the axis of trap 1. After three cleaning cycles of typically 2 seconds, all dark ions are lost out of the trap. The \Mg ions have a survival probability of \textgreater 95\%. A pure \Mg ion crystal after successful isotope selective loading is shown in figure \ref{fig:ionenbild} a).

\begin{figure}
	\includegraphics[width=0.5\textwidth]{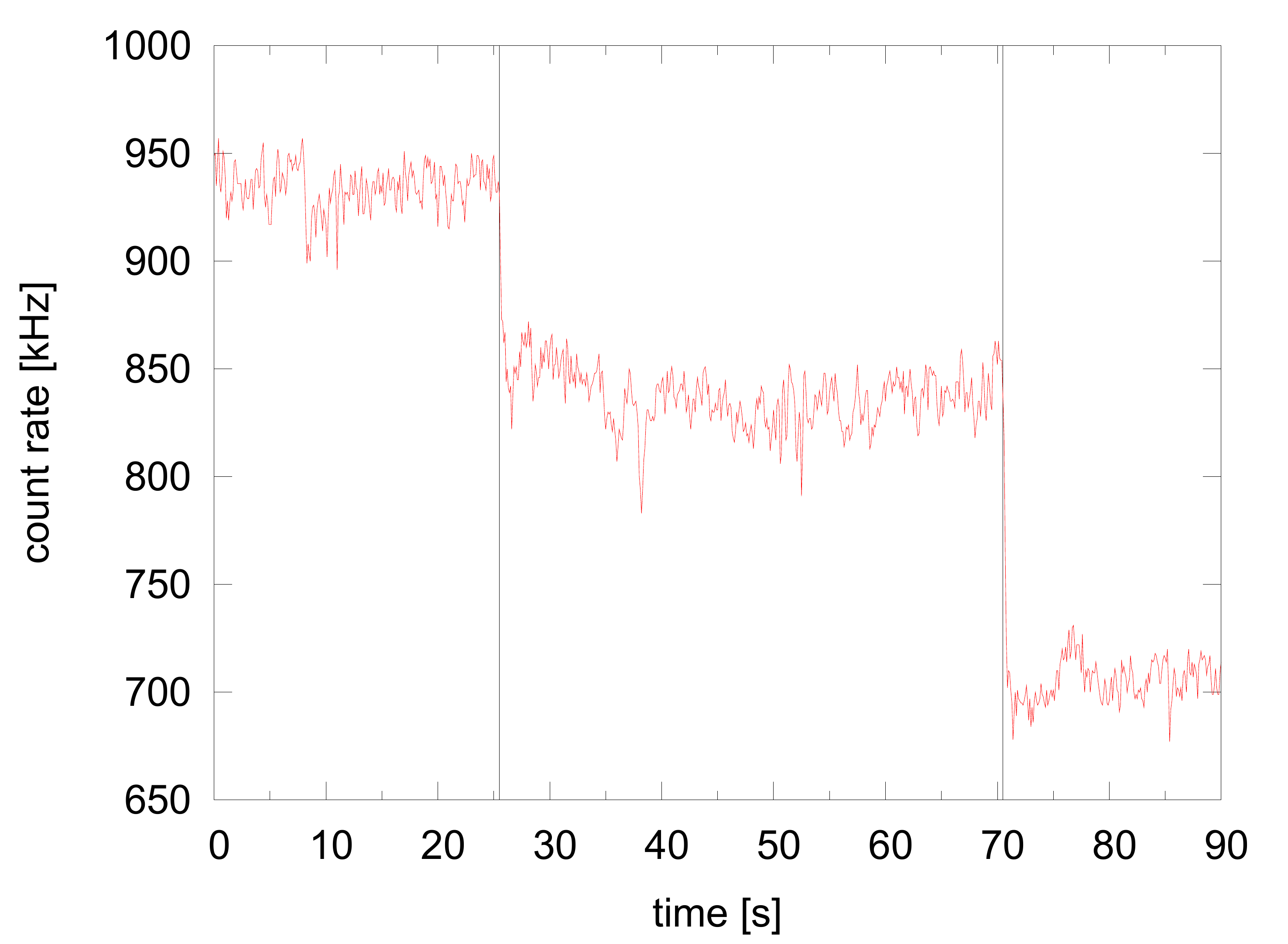}
	\caption{Breading of molecular ions via the photochemical reaction (\ref{reaction}): The count rate, detected by a photomultiplier tube is proportional to the amount of \Mg ions in trap 1. Hydrogen injection was started at the indicated instants (vertical lines). The count rate drops twice when $15\pm 3\,\%$ and $25\pm 3\,\%$ of the atomic ions are converted into \MgH. The background count rate (no ions in the trap) amounts to 320\,kHz. Each data point represents an accumulation time of 100\,ms.}
	\label{fig:PMTCountRate}
\end{figure}

After the preparation of a pure atomic \Mg crystal, small amounts of hydrogen gas are injected into chamber 1. Simultaneously, the atomic ions are exposed to 1.5\, mW of the near resonant (about $\Gamma/2$ red detuned) cooling light, focused down to 80\,\micro m beam waist, leading to a significant population of the $\text{3p}\,\text{P}_{3/2}$ state and therefore to approximately 4\,eV of additional energy allowing for the photo-chemical reaction (see equation \ref{reaction}). This provides a controlled transformation of a selectable amount of ions in the crystal to the molecular ion model system \MgH. Opening the pulsed valve for 5-15\,ms with the hydrogen reservoir at a pressure of $\approx$0.3\,mbar, rising the pressure to a few $10^{-9}$\,mbar for a period of less than 2 seconds, yields a conversion of  nearly 50\,\% of the \Mg ions into \MgH. Figure \ref{fig:PMTCountRate} shows the count rate of the PMT, detecting the fluorescence light of atomic ions in trap 1 and its change due to two injections of hydrogen. After opening the valve, the rate drops by 20\% due to the intended conversion of a fraction of the \Mg ions into non-fluorescing molecular ions. After about 250 milliseconds the reactions cease indicating fast pumping of the injected hydrogen gas. A typical bi-crystal, breaded as described above is shown in figure \ref{fig:ionenbild} d).

\begin{figure}
	\includegraphics[width=0.5\textwidth]{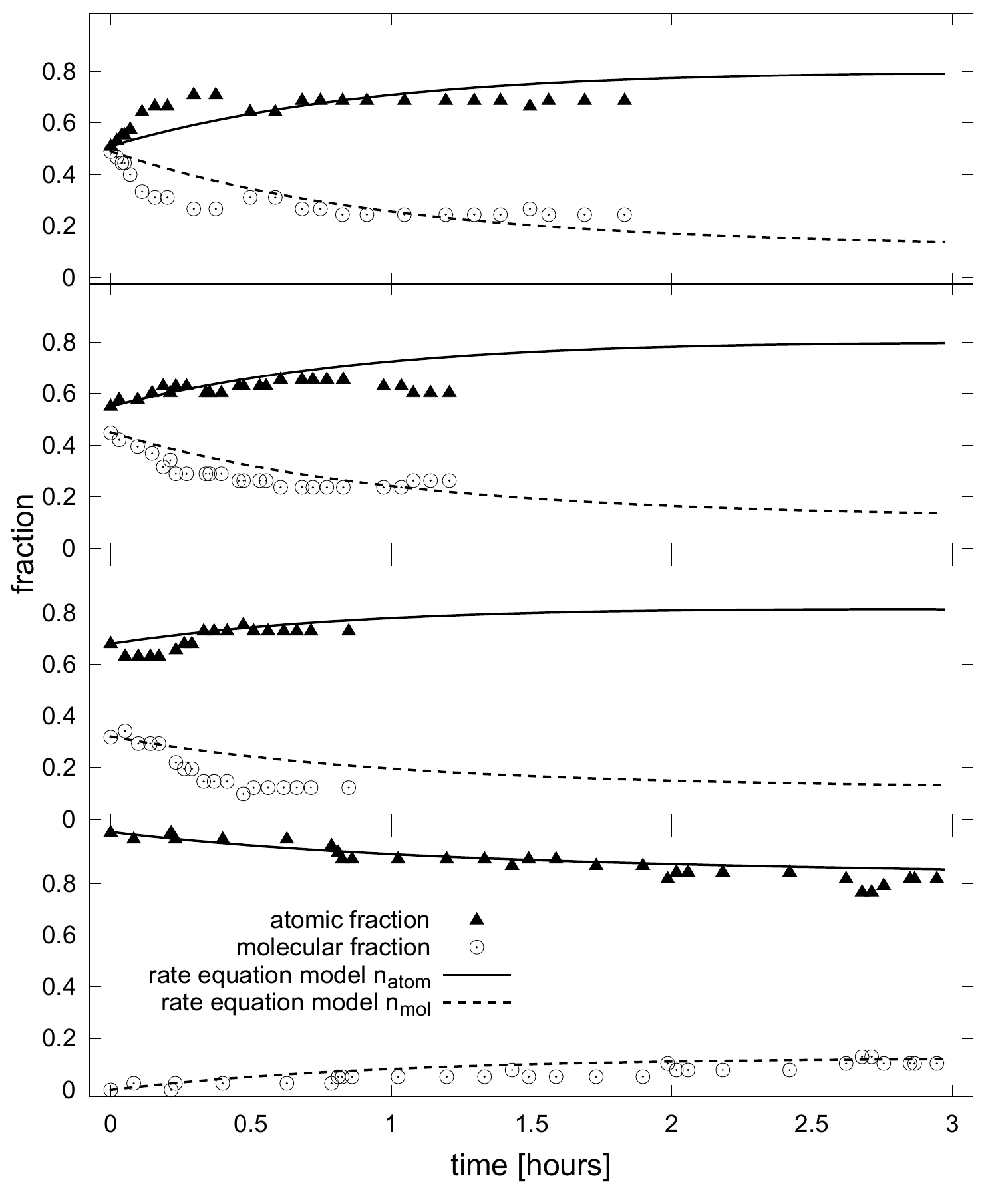}
	\caption{Atomic \Mg and molecular \MgH ion composition of four Coulomb crystals in dependence of the storage duration in trap 2. The atomic fraction of the initial crystal increases from 50\% to 100\% for the experiments from top to bottom. The compositions are derived out of fluorescence images, taken by the EM-CCD camera. The laser power was chosen sufficient to sustain the crystalline state of the ions in the trap ($\approx150\,mW/cm^2$ intensity and about one natural linewidth detuning). According to a nude Bayard-Alpert type ionization gauge (calibrated on $\text{N}_2$), the total pressure of the vacuum chamber amounts to approximately $2\cdot10^{-10}$ mbar. The results derived with a rate equation model of the time evolution of the atomic and molecular fraction are shown as solid and dashed lines. The evolution of all four ion-crystals is described with a single set of rates (see main text). No counting error in the evaluation of fluorescence images is assumed, therefore the composition of each of the observed crystals can be given exactly.}
	\label{fig:backgroundReactions}
\end{figure}

In order to investigate the requirement to separate the region for production and the precision experiment in the setup, the rates of laser-induced reactions of the \MgH model system have been studied, similar to reference \cite{molhave00}. In figure \ref{fig:backgroundReactions}, the time dependent composition of two-component ion crystals with different fractions of \Mg and \MgH are presented. The experiment was performed under UHV conditions over an extended period of time (up to 3 hours). The rates of the reaction of photodissoziation described in equation (\ref{reaction}) are proportional to the population of the excited state $\text{3p}\,\text{P}_{3/2}$. Thus, in order to minimize the rate of reactions in this study, the laser power of the cooling beam was set to just sustain the crystalline structure. A rate equation model was developed to describe the time evolution of the fraction of atomic to molecular ions. Three light induced reactions are considered and reaction rate coefficients $\kappa_1, \kappa_2$ and $\kappa_3$ were fitted to each process. One is the photochemical reaction described in equation \ref{reaction} ($\kappa_1=0.13\,\text{h}^{-1}$) and the other two are the dissociation reactions $\MgH\rightarrow\Mg+\text{H}$ ($\kappa_2=0.79\,\text{h}^{-1}$)
and $\MgH\rightarrow \text{\textsuperscript{24}Mg} + \text{H\textsuperscript{+}}$ ($\kappa_3=0.10\,\text{h}^{-1}$). This set of rates reproduces the measured time dependent composition of the four experiments and the time scale for the relevant reactions is consistent with results reported in \cite{staanum10}. The limited statistics gained within the experiments can not provide an accurate measurement of the reaction rates, however, it yields a rough estimate of the relevant time scales for the specific partial pressure of the reactional gases within chamber number 3. To conclude, stable conditions for Coulomb-crystals that allow for precise control of the position of individual ions require ultra-high vacuum conditions (\textless$10^{-10}$\,mbar). Experiments with molecular ions, even with the reactive molecular ion \MgH, are possible on the time scale of minutes when performed in the UHV. These conditions are difficult to provide, if the production and delivery rate of the molecular ion and its surrounding Coulomb-crystal exceeds 1/s or has to be achieved in a continuous way for example by Electrospray Ionization. For these applications, a deterministic transfer of individual molecular ions from the production to a well isolated experimental region is required and becomes a powerful tool. For example, the transfer schemes presented in the next section are advantageous when an implementation of buffer gas cooling of the molecular ions or a deterministic initialization and a fast replacement of single molecular ions is required.  

\subsection{Deterministic transfer}
For the purpose of demonstrating the deterministic transfer scheme, a molecular ion is prepared and confined in trap 1. The ion is accelerated and the quadrupole guides it through chamber 2 and the two differential pumping stages, described in section \ref{the experimental setup}. The quadrupole ends in the third chamber of the experimental apparatus, that maintains UHV conditions. The transfered ion is then decelerated and subsequently trapped by the Paul traps 2 and/or 3 located at the end of the guide. The three ring electrodes, that form the axial confining potential for these two traps are referred to as $\text{R}_1$, $\text{R}_2$ and $\text{R}_3$ in the following (see figure \ref{fig:apparat}B).

\begin{figure}
	\centering
		\includegraphics[width=0.50\textwidth]{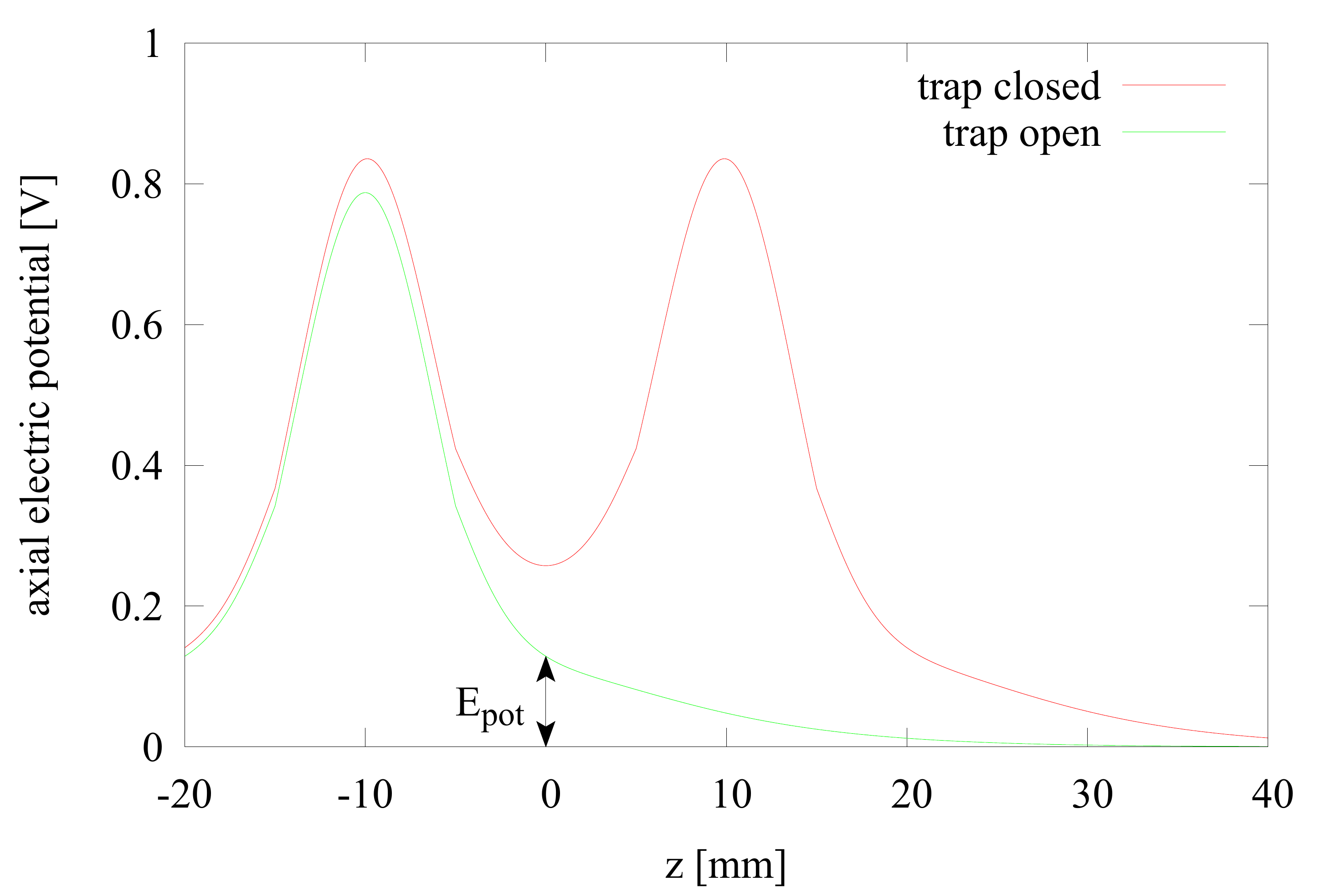}
	\caption{Results of a simulation of the axial electric potential provided by voltages applied to the ring electrodes of trap 1 taking the shielding effect of the rf-electrodes into account. The result achieved for applying 250\,V to both ring-electrodes is depicted by the red line. The trap is opened axially by switching one of the ring-electrodes to electrical ground (green line). The ions are accelerated along the RF-guide by the potential gradient and their excess potential energy $\text{E}_\text{pot}$ is converted into kinetic energy.}
	\label{fig:beschleunigung}
\end{figure}

\begin{figure*}
	\includegraphics[width=0.5\textwidth]{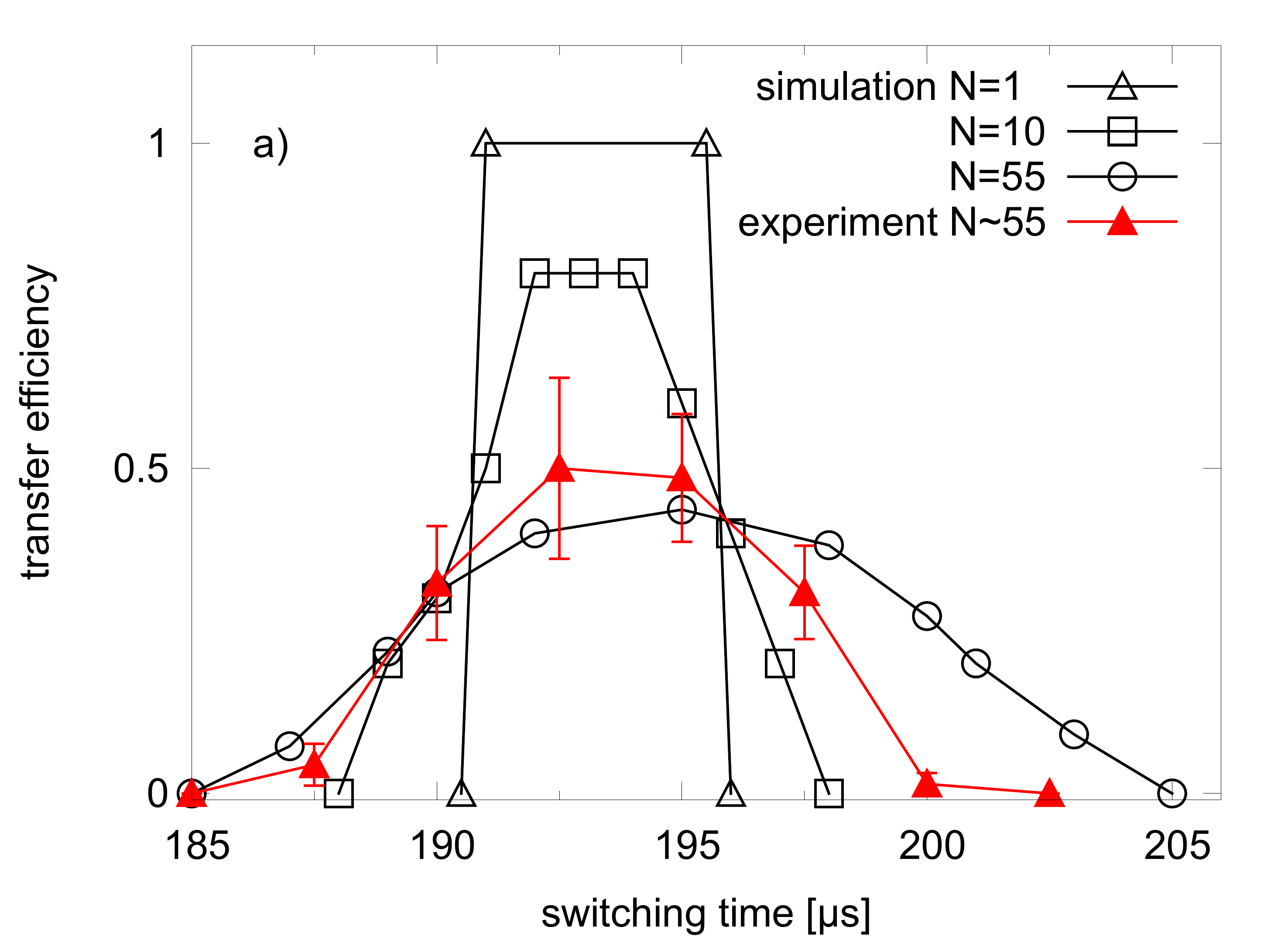}
	\includegraphics[width=0.5\textwidth]{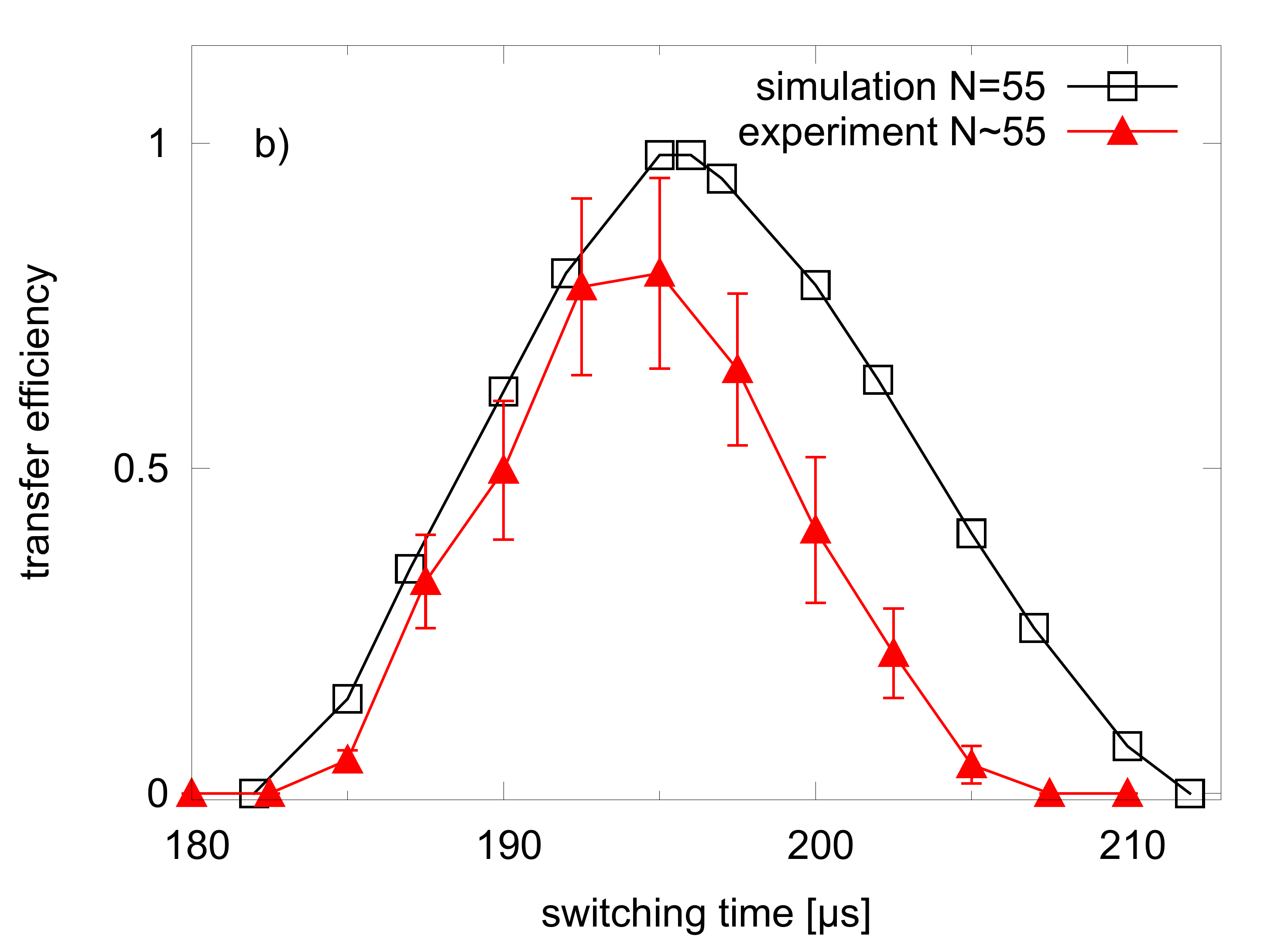}
	\includegraphics[width=0.5\textwidth]{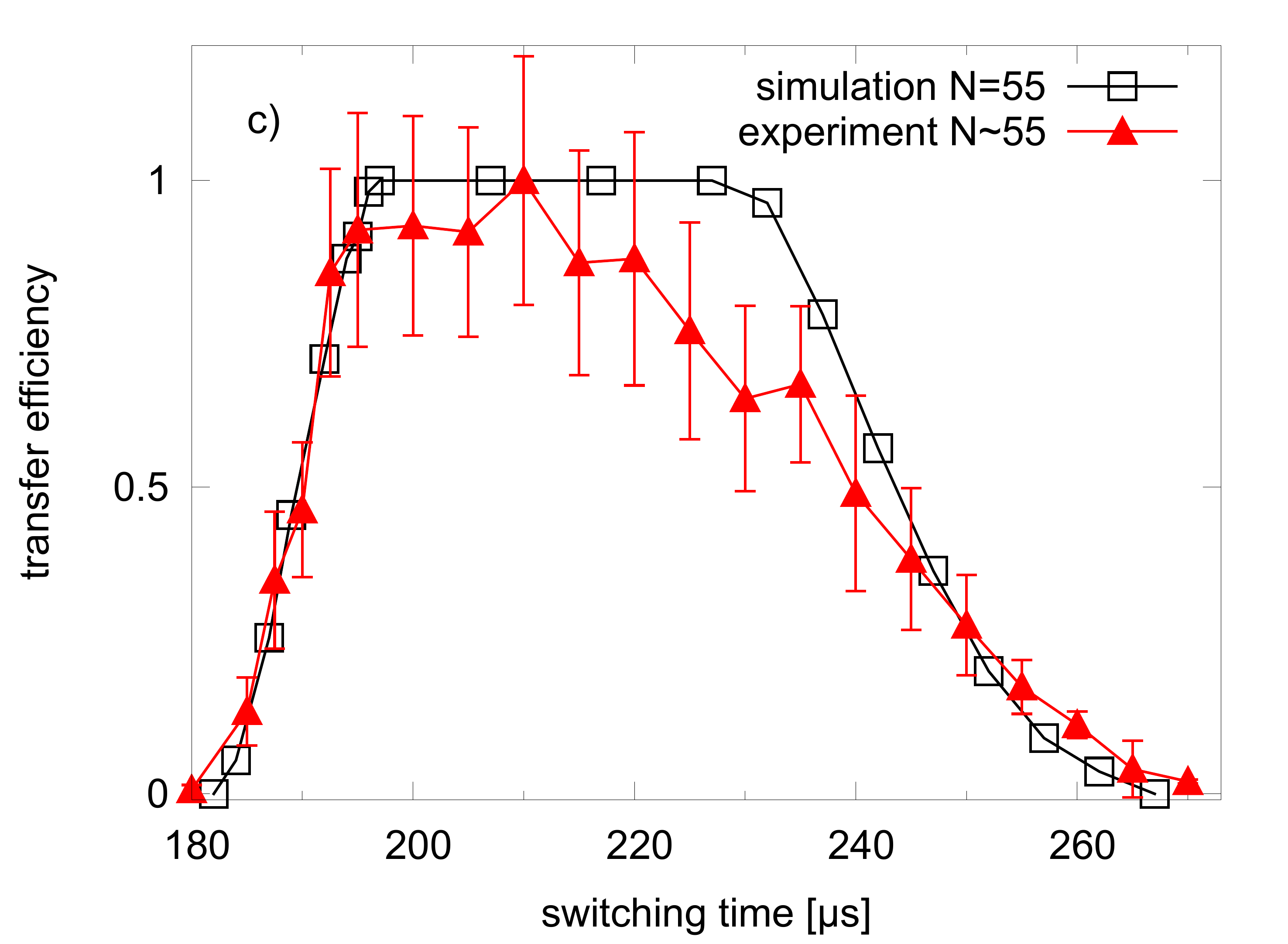}
	\caption{Measured and numerically simulated transfer efficiencies of an ensemble of ions from trap 1 over a distance of 39\,cm to trap 2 in dependence of the delay of switching the involved ring electrodes (time of flight of the ensemble). Figure a) shows the results of simulations of the transfer efficiency of method 1 for different initial particle numbers. The measurements were performed under conditions comparable to those assumed in the N=55 simulation. Figure b) and figure c) compare the experimental deduced data with the simulation results for 55 ions achieved following transfer method 2 and 3, respectively. The presented error bars take the uncertainties of the initial number of ions in trap 1 into account. The fluorescence count rate of the photomultiplier in chamber 1, used to load initial crystals with similar particle numbers, yields an estimated uncertainty of $\pm$ 15\,\% for the absolute scale of the measurements. Furthermore, discrepancies between simulation and experiment are due to uncertainties about the exact shape of the axial electric potential of the traps 1 and 2. Additional voltages on the micro-motion compensation electrodes cause deviations from the idealized axial potential solely gernerated by the ring-electrodes. Nevertheless, trends of the simulation, the different durations of sucessful re-capture and the different absolute efficiencies are in agreement with the experimental results.}
	\label{fig:shuttelntrap2}
\end{figure*}

The molecular ion is released out of the axial confinement of trap 1 by grounding the ring electrode facing the differential pumping stage (see figure \ref{fig:apparat}A). Switching of the voltages by MOS-FET transistors takes currently less than 1\,\micro s. The ion is accelerated along the axis of the guide by applying 250\,V to the first ring-electrode of trap 1 (see figure \ref{fig:beschleunigung}). A CPO\footnote{Charged Particle Optics programs, CPO Ltd} simulation reveals a substantial shielding effect caused by the quadru\-pole electrodes, related to a reduced electric axial potential at the trap center by approximately three orders of magnitude. The remaining excess potential energy of 130\,meV is converted into axial kinetic energy of the ion. Time of flight measurements reveal an average ion velocity of 1850\,ms\textsuperscript{-1}, indicating that the voltages applied on micro-motion compensation wires near trap 1 contributed to the acceleration. After approximately 190\,\micro s of propagation without axial confinement, the molecular ion arrives at the optional Paul traps 2 and 3. Approaching the center position between the two ring electrodes $\text{R}_1$ and $\text{R}_2$, the voltages are applied and the guided, transfered ion is trapped. For a single molecular ion, this transfer can be processed with nearly 100\% efficiency, because the radial confining RF field provides a deep trapping potential (typical 1\,eV) rendering radial losses of ions negligible. Simulations of the axial dynamic of the particle in the time-depending potentials of the transfer process by numerically solving Newton's equation of motion predict a 5\,\micro s (see figure \ref{fig:shuttelntrap2} a)) time window, where a single ion can be trapped reliably using the protocol and parameters described above.  

If required, the transfer of tens to hundreds of individual ions is possible at once with the discussed scheme. However, the efficiency suffers for an increased number of ions. The mutual Coulomb repulsion in an ensemble, launched out of trap 1, causes an axial spreading of the ensemble during propagation. The maximal number of ions that can be decelerated and trapped, depends on the distance between the ring-electrodes, the average time of flight and the parameters for axial confinement in trap 1. Stiffer axial confinement is related to a higher potential energy, that is converted into kinetic energy and thus leads to a shorter time of flight. On the other hand, the related reduced inter-ion spacing in trap 1 leads to an initially increased Coulomb repulsion and hence to a faster spreading of the ensemble during the flight. Within the chosen range of parameters the influence of the shorter time of flight dominates and the increased Coulomb repulsion has only an influence on the order of a few percent on the maximal number of ions that can be trapped in total.

\begin{figure}
	\centering
		\includegraphics[width=0.5\textwidth]{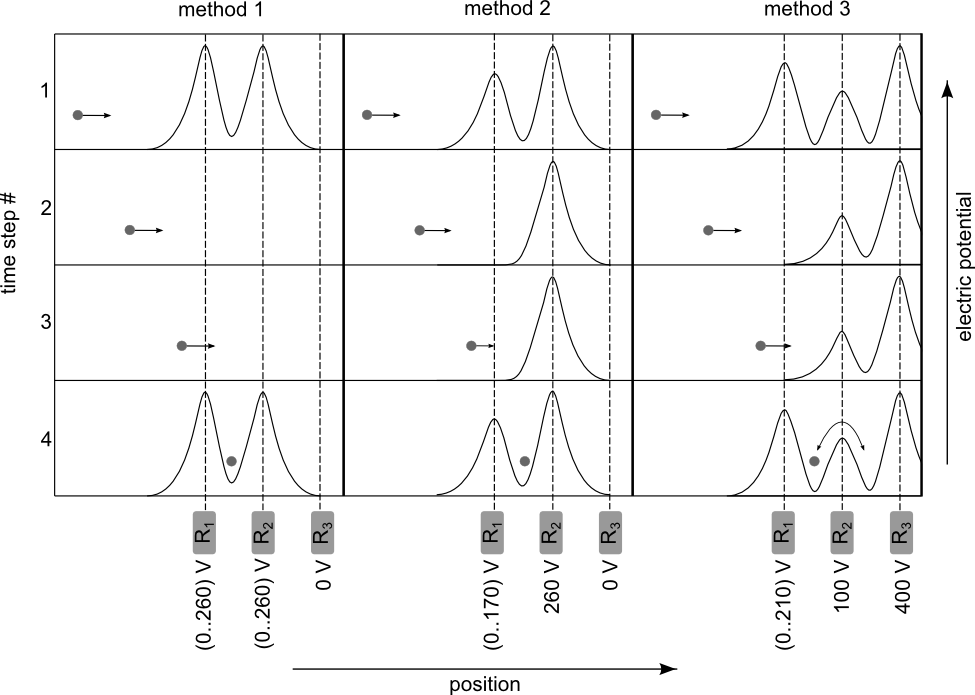}
	\caption{Schematic overview of three different methods used to decelerate and trap various numbers of ions at the end of the transfer process. Different timings and amplitudes of voltages on the ring-electrodes $\text{R}_1$, $\text{R}_2$ and $\text{R}_3$ (see figure \ref{fig:apparat}) provide the switching of the axial electric potential along the radially confining RF-guide. The dot and arrow represent a single or the center of mass of a bunch of ions and their average velocity, respectively. The position relative to the switching axial potential wells is shown for different time steps.}
	\label{fig:shuttelnschematisch}
\end{figure}

To enhance the transfer efficiency for an increasing number of ions, three different methods (see figure \ref{fig:shuttelnschematisch}) have been developed. The first method is closely related to the single ion transfer scheme discussed above. The ring electrode $\text{R}_3$ is grounded and $\text{R}_1$ and $\text{R}_2$ are switched simultaneously between 0\,V and 260\,V. The switching is synchronized with the release of the crystal from trap 1, in order to close trap 2 when the maximal fraction of ions can be enclosed. The transfer efficiency of method 1 for an ion crystal, that contains 10 ions, is reduced to 80\%, for 50 ions it drops to 50\%. Figure \ref{fig:shuttelntrap2} a) shows the result of simulations and of a measurement on the transfer efficieny in dependence of the time delay between launching the ions and switching of electrodes $\text{R}_1$ and $\text{R}_2$. Experimentally, the initial number of ions in trap 1 is derived from the averaged photon count rate of the photomultiplier. The absolute number of ions and the measurement of the absolute transfer efficiency are accurate to $\pm$15\%. To increase the accuracy, a second CCD camera is necessary. 

A higher efficiency to decelerate and trap a spreading ion ensemble can be achieved via method 2 by switching only the ring electrode $\text{R}_1$ from ground to 170\,V leaving the voltage on $\text{R}_2$ and $\text{R}_3$ constant. The voltage applied to $\text{R}_2$ decelerates the ensemble and finally reflects it. This leads to a double-pass and a bunching of the ensemble in the trapping region, increasing the ion density and thus extending the time window for trapping compared to method 1. For the given parameters, the largest ion crystal, that can be theoretically recaptured without losses contains 55 ions. Figure \ref{fig:shuttelntrap2} b) compares the simulated transfer efficiency for initially 55 ions with the results of the performed experiments. While method 1 allowed 50\% transfer efficiency, 80\% is reached using method 2.

The third method uses the ring electrode $\text{R}_3$ as an additional reflector for ions having a kinetic energy allowing to surpass $\text{R}_2$. The potential generated by $\text{R}_1$ is raised sufficiently high to prevent reflected ions from escaping in backward direction. The voltages on $\text{R}_2$ and $\text{R}_3$ were set to 100\,V and 400\,V, respectively and kept constant during the whole transfer procedure. $\text{R}_1$ is switched from 0\,V to 210\,V. The asymmetry between the voltages on $\text{R}_1$ and $\text{R}_3$ provides an axial positioning of the ions where micro-motion compensation is optimal. The fraction of the ensemble that overcomes the potential hill generated by the middle ring electrode starts oscillating between the two potential minima. However, laser cooling reduces the kinetic energy of the \Mg-ions and the exchange between the two Paul traps ceases on time scales of typically a few seconds. The duration required to re-crystallize the ensemble of atomic and molecular ions critically depends on the laser intensity, the detuning from the atomic resonance and sufficient access of all motional degrees of freedom. Currently, one axial cooling beam enclosing only a small (few degrees) angle with the guide axis, barely addresses the radial degree of freedom. Therefore, mainly the normal modes of the crystal at a linear density allowing for zigzag and higher dimensional structures sufficiently couple the longitudinal and radial motion. A laser beam, that is red-detuned by about a natural linewidth of the atomic transition, as it is required to reach low temperatures for a crystal, is far off-resonant due to the Doppler shift for ions moving at velocities $\gg$1\,m/s. A different spatial overlap of the cooling beam with the ion crystals in trap 2 and 3 due to an imperfect matching of the two axis, leads to different final distributions of ions in the two traps. Figure \ref{fig:shuttelntrap2} c) shows the transfer efficiency into trap 2 as a function of the time delay for switching $\text{R}_1$ using method 3. The experiment qualitatively confirms the duration for maximizing the probability for recapture and the asymmetric shape for long and short switching times predicted by the simulation.

Using the transfer method 3, the traps 2 and 3 get both populated with ions. An interchange of the ions between the traps can be performed at an efficiency close to 100\%. The two traps share the electrode $\text{R}_2$ and the tight axial confinement is preserved for all times. Switching electrode $\text{R}_2$ to electrical ground for 25\,\micro s leads to an oscillation of the two ensembles centered at the new potential minimum. No change in the fluorescence rate on the CCD camera was detected when interchanging the ensembles of ions, indicating a unit efficiency of the transfer and that the crystalline structure is preserved during the oscillation (if unlikely a melting did occurr, the re-crystallization happened too fast to be detected by the CCD with a 100\,ms exposure time). In order to analyze the potential interaction of, or exchange between the two initial ensembles, a distinguishable composition of molecular and atomic ions were prepared in trap 2 and 3. If the duration (25\,\micro s) of grounding $\text{R}_2$ coincidences with half the oscillation period of the ions, a complete exchange of the ensembles between trap 2 and 3 is observed. A redistribution of the total amount and composition ratio of ions in the two traps is achieved by using a grounding duration different from the 25\,\micro s. For example, a random distribution of the ions is achieved by reducing the grounding duration and a gathering of $\approx$ 90\% of the ions into one of the trap is achieved by increasing the grounding duration.

\section{Conclusions and outlook} 
We have shown the deterministic preparation of a selectable amount of externally cold and well localized molecular ions down to a single molecular ion. A pure \Mg ion crystal was loaded into a trap at the front end of a quadrupole ion guide. The model system \MgH \newline was produced by the photochemical reaction of \Mg \newline with hydrogen gas. The resulting molecular ion can be transfered along the ion guide and re-trapped by the second or third Paul trap at the end of the guide with unit efficiency. Three different transfer schemes for a mixed atomic and molecular ion ensemble and its corresponding transfer efficiencies have been investigated. The demonstrated highly efficient exchange of ion ensembles trapped in two separated traps can be used to load an ion trap with a transfered ion ensemble serving as a reservoir.

The method presented in this paper can be extended to other molecular ions using electron bombardment ionization or the already implemented Electrospray Ionization source (ESI) \cite{fenn89} combined with an additional charge-to-mass ratio filtering stage. Together with directly laser cooled \textsuperscript{138}\text{Ba}\textsuperscript{+} ions (already implemented in the presented setup) \cite{Bapapier}, protonated molecules out of the ESI-source cover a wide range of charge-to-mass up to biologically relevant molecular ions. The presented transfer schemes can be enhanced in efficiency by using the electrodes inside the differential pumping stages as drift tubes to additionally bunch an ensemble during transfer \cite{schramm01} or to even realize a transfer where the crystalline structure of the ions is sustained \cite{schramm02}, significantly improving the re-cooling times in the destination trap. An additional cooling of the internal degrees of freedom of the molecular ions by lasers \cite{staanum10,schneiderschiller10} or by buffer gas \cite{gerlich09} is realizable for example in chamber 1 or 2. The cooling light pressure, that acts on atomic ions only, can be used to separate the molecular from the atomic fraction of the crystal. Extending the setup by an additional segmented trap will give the possibility to cut single molecular ions out of a heterogeneous crystal and will be continuously loaded by exploiting the presented exchange of ensembles between two traps. The separation of individual molecular ions out of a two component crystal has already been demonstrated by our group, using a different segmented Paul-trap setup \cite{schaetz07}. The presented production and control scheme for cold molecular ions can serve as a source for container-free molecular targets, in principle coolable to the motional ground state \cite{barrett03}, for imaging or spectroscopic analysis \cite{mghpapier}.

\begin{acknowledgements}
The authors would like to thank W. Schmid, T. Hasegawa, J. Bayerl, C. Kerzl and T. Dou for their contributions and W. Fuss for fruitful discussions and his encouragement. Financial support is gratefully acknowledged by the Deutsche Forschungsgemeinschaft, the DFG cluster of Excellence: Munich Center for Advanced Photonics, the International Max Planck Research School on Advanced Photon Science (IMPRS-APS) and the EU research project PICC: The Physics of Ion Coulomb Crystals, funded under the European Community's 7th Framework Programme.
\end{acknowledgements}

\bibliography{referenzen}

\begin{thebibliography}{10}

\bibitem{Bethlem2000}
H.~L. Bethlem, {\it et~al.\/}, {\it Nature\/} {\bf 406}, 491 (2000).

\bibitem{junglen04b}
T.~Junglen, T.~Rieger, S.~A. Rangwala, P.~W.~H. Pinkse, G.~Rempe, {\it Phys.
  Rev. Lett.\/} {\bf 92}, 223001 (2004).

\bibitem{Bethlem1999}
H.~L. Bethlem, G.~Berden, G.~Meijer, {\it Phys. Rev. Lett.\/} {\bf 83}, 1558
  (1999).

\bibitem{rangwala03}
S.~A. Rangwala, T.~Junglen, T.~Rieger, P.~W.~H. Pinkse, G.~Rempe, {\it Phys.
  Rev. A\/} {\bf 67}, 043406 (2003).

\bibitem{molhave00}
K.~M\o{}lhave, M.~Drewsen, {\it Phys. Rev. A\/} {\bf 62}, 011401 (2000).

\bibitem{Roth2005}
B.~Roth, A.~Ostendorf, H.~Wenz, S.~Schiller, {\it Journal of Physics B: Atomic,
  Molecular and Optical Physics\/} {\bf 38}, 3673 (2005).

\bibitem{neutze00}
R.~Neutze, R.~Wouts, D.~van~der Spoel, E.~Weckert, J.~Hajdu, {\it Nature\/}
  {\bf 406}, 752 (2000).

\bibitem{chapman06}
H.~N. Chapman, {\it et~al.\/}, {\it Nature Physics\/} {\bf 2}, 839 (2006).

\bibitem{webster02}
G.~Webster, R.~Hilgenfeld, {\it Single Molecules\/} {\bf 3}, 63 (2002).

\bibitem{henderson95}
R.~Henderson, {\it Quarterly Reviews of Biophysics\/} {\bf 28}, 171 (1995).

\bibitem{young10}
L.~Young, {\it et~al.\/}, {\it Nature\/} {\bf 466}, 56 (2010).

\bibitem{seibert11}
M.~Seibert, {\it et~al.\/}, {\it Nature\/} {\bf 470}, 78 (2011).

\bibitem{barrett03}
M.~D. Barrett, {\it et~al.\/}, {\it Phys. Rev. A\/} {\bf 68}, 042302 (2003).

\bibitem{gerlich09}
D.~Gerlich, G.~Borodi, {\it Faraday Discussions\/} {\bf 142}, 57 (2009).

\bibitem{Paul90}
W.~Paul, {\it Rev. Mod. Phys.\/} {\bf 62}, 531 (1990).

\bibitem{willitsch08}
S.~Willitsch, M.~T. Bell, A.~D. Gingell, S.~R. Procter, T.~P. Softley, {\it
  Physical Review Letters\/} {\bf 100}, 043203 (2008).

\bibitem{kjaegaard00}
N.~Kjaergaard, L.~Hornekaer, A.~Thommesen, Z.~Videsen, M.~Drewsen, {\it Applied
  Physics B: Lasers and Optics\/} {\bf 71}, 207 (2000).

\bibitem{rotter08}
D.~Rotter, Quantum feedback and quantum correlation measurements with a single
  barium ion, Ph.D. thesis (2008).

\bibitem{Bapapier}
G.~Leschhorn, T.~Hasegawa, T.~Schaetz, {\it ArXiv e-prints:1110.4040\/}
  (2011).

\bibitem{madsen00}
D.~N. Madsen, {\it et~al.\/}, {\it Journal of Physics B: Atomic, Molecular and
  Optical Physics\/} {\bf 33}, 4981 (2000).

\bibitem{diedrich87b}
F.~Diedrich, E.~Peik, J.~M. Chen, W.~Quint, H.~Walther, {\it Physical Review
  Letters\/} {\bf 59}, 2931 (1987).

\bibitem{wineland87}
D.~J. Wineland, J.~C. Bergquist, W.~M. Itano, J.~J. Bollinger, C.~H. Manney,
  {\it Physical Review Letters\/} {\bf 59}, 2935 (1987).

\bibitem{friedenauer06}
A.~Friedenauer, {\it et~al.\/}, {\it Applied Physics B: Lasers and Optics\/}
  {\bf 84}, 371 (2006).

\bibitem{smith71}
P.~W. Smith, T.~W. H\"ansch, {\it Phys. Rev. Lett.\/} {\bf 26}, 740 (1971).

\bibitem{wieman76}
C.~Wieman, T.~W. H\"ansch, {\it Phys. Rev. Lett.\/} {\bf 36}, 1170 (1976).

\bibitem{fenn89}
J.~B. Fenn, M.~Mann, C.~K. Meng, S.~F. Wong, C.~M. Whitehouse, {\it Science\/}
  {\bf 246}, 64 (1989).

\bibitem{ostendorf06}
A.~Ostendorf, {\it et~al.\/}, {\it Physical Review Letters\/} {\bf 97}, 243005
  (2006).

\bibitem{larson86}
D.~J. Larson, J.~C. Bergquist, J.~J. Bollinger, W.~M. Itano, D.~J. Wineland,
  {\it Phys. Rev. Lett.\/} {\bf 57}, 70 (1986).

\bibitem{Drewsen04}
M.~Drewsen, A.~Mortensen, R.~Martinussen, P.~Staanum, J.~L. S\o{}rensen, {\it
  Phys. Rev. Lett.\/} {\bf 93}, 243201 (2004).

\bibitem{staanum10}
P.~F. Staanum, K.~Hojbjerre, P.~S. Skyt, A.~K. Hansen, M.~Drewsen, {\it Nature
  Physics\/} {\bf 6}, 271 (2010).

\bibitem{schramm01}
U.~Schramm, T.~Schaetz, D.~Habs, {\it Phys. Rev. Lett.\/} {\bf 87}, 184801
  (2001).

\bibitem{schramm02}
U.~Schramm, T.~Schaetz, D.~Habs, {\it Phys. Rev. E\/} {\bf 66}, 036501 (2002).

\bibitem{schneiderschiller10}
T.~Schneider, B.~Roth, H.~Duncker, I.~Ernsting, S.~Schiller, {\it Nature
  Physics\/} {\bf 6}, 275 (2010).

\bibitem{schaetz07}
T.~Schaetz, A.~Friedenauer, H.~Schmitz, L.~Petersen, S.~Kahra, {\it Journal of
  Modern Optics\/} {\bf 54}, 2317 (2007).

\bibitem{mghpapier}
S.~Kahra, G.~Leschhorn. Controlled delivery of single molecules into
  ultra-short laser pulses: a molecular conveyor belt, (submitted 2011).

\end{thebibliography}
\bibliographystyle{science}  

\end{document}